\documentclass[acmtog,nonacm]{acmart}

\acmSubmissionID{1693}

\usepackage{booktabs} 

\usepackage[ruled]{algorithm2e} 

\SetAlFnt{\small}
\SetAlCapFnt{\small}
\SetAlCapNameFnt{\small}
\SetAlCapHSkip{0pt}

\usepackage{subcaption}
\usepackage{svg}

\setcopyright{cc}

\begin{document}
\title{Haptic Tracing: A new paradigm for spatialized Haptic rendering}


\author{Tom Roy}
\affiliation{
  \institution{Inria, Univ. Rennes, CNRS, IRISA}
  \city{Rennes}
  \country{France}}
\email{tom.roy@inria.fr}
\author{Yann Glemarec}
\affiliation{
  \institution{Inria, Univ. Rennes, CNRS, IRISA}
  \city{Rennes}
  \country{France}}
\email{yann.glemarec@inria.fr}
\author{Gurvan Lecuyer}
\affiliation{
  \institution{Interdigital}
  \city{Cesson-Sevigné}
  \country{France}}
\email{gurvan.lecuyer@interdigital.com}
\author{Quentin Galvane}
\affiliation{
  \institution{Interdigital}
  \city{Cesson-Sevigné}
  \country{France}}
\email{quentin.galvane@interdigital.com}
\author{Philippe Guillotel}
\affiliation{
  \institution{Interdigital}
  \city{Cesson-Sevigné}
  \country{France}}
\email{philippe.guillotel@interdigital.com}
\author{Ferran Argelaguet}
\affiliation{
  \institution{Inria, Univ. Rennes, CNRS, IRISA}
  \city{Rennes}
  \country{France}}
\email{ferran.argelaguet@inria.fr}

\newcommand{\CR}[1]{{\color{red}{#1}}}
\newcommand{\tom}[1]{\mycomment{author1}{TR}{#1}}
\newcommand{\yann}[1]{\mycomment{author2}{YG}{#1}}
\newcommand{\quentin}[1]{\mycomment{author3}{QG}{#1}}
\newcommand{\gurvan}[1]{\mycomment{author4}{GL}{#1}}
\newcommand{\ferran}[1]{\mycomment{author5}{FA}{#1}}

\begin{abstract}




Haptic technology enhances interactive experiences by providing force and tactile feedback, improving user performance and immersion. However, despite advancements, creating tactile experiences still remains challenging due to device diversity and complexity. Most available haptic frameworks rely on trigger-based or event-based systems, and disregard the information of the 3D scene to render haptic information. This paper introduces Haptic Tracing, a novel method for spatial haptic rendering that simplifies the creation of interactive haptic experiences without relying on physical simulations. It uses concepts from visual and audio rendering to model and propagate haptic information through a 3D scene.
The paper also describes how our proposed haptic rendering method can be used to create a vibrotactile rendering system, enabling the creation of perceptually coherent and dynamic haptic interactions.
Finally, the paper discusses a user study that explores the role of the haptic propagation and multi-actuator rendering on the users' haptic experience. 
The results show that our approach significantly enhances the realism and the expressivity of the haptic feedback, showcasing its potential for developing more complex and realistic haptic experiences.

\end{abstract}

%
%


%
%


\maketitle
\section{Introduction}

Haptic technology refers to systems capable of eliciting force (muscles and tendons) and tactile sensations (mechanoreceptors in the skin), including force feedback, thermal, vibratory or even electrotactile devices~\cite{culbertson2018haptics}.
Haptic feedback, in addition to improve motor tasks and learning~\cite{sullivan2021haptic}, can increase the sense of presence and embodiment in interactive 3D experiences~\cite{slater2010first}.
Over the past decades, research in this field has been highly prolific, leading to the development of a wide range of new rendering devices, frameworks, and applications across diverse domains, specially on haptic wearable devices~\cite{Pacchierotti17}. Haptic devices are little by little penetrating the consumer market (e.g., vests, gloves and headphones), driving standardization efforts~\cite{guillotel2025adding}.
However, while physical simulation is the gold standard for kinesthetic feedback devices~\cite{laycock2007survey}, the crafting of interactive experiences with tactile feedback devices remains challenging. Tactile systems are extremely heterogeneous~\cite{Pacchierotti17}, can be spatialized over user's body~\cite{Danieau2018, Yun2023}, and use hand-crafted haptic effects~\cite{weber2011evaluation,Schneider2016,terenti2023vireo}. 
%
This heterogeneity has primarily steered research toward device prototyping, perceptual studies, and haptic signal authoring. Progress in these areas contrasts sharply with the relative simplicity of haptic authoring tools for interactive experiences. Most existing tools remain trigger-based or event-based~\cite{Interhaptics}, and totally disregard the spatial structure of the 3D environment when modulating haptic information.

In this paper we propose Haptic Tracing, a novel spatial haptic rendering method to ease the creation of interactive and spatialized haptic experiences with a focus on 3D interactive scenarios.
The goal of the proposed method is to enable the creation of rich haptic-enhanced experiences without increasing the authoring/development cost, ensuring a real-time haptic rendering and maximizing the user haptic experience~\cite{kim2020defining}.
The proposed method, compared to traditional force-feedback haptic rendering methods, does not rely on physical simulation methods~\cite{HapticRenderingDef}, but considers that haptic effects are synthetic, and crafted by designers with a specific artistic intent.
We leverage concepts from visual (i.e., ray tracing rendering) and audio rendering (e.g., spatial audio rendering) to propose a data structure and a haptic rendering algorithm to model a haptic scene, define how haptic information is propagated, and how it is ultimately rendered within a dedicated device. 
As several aspects of the rendering method are dependent on the specificity of the haptic device/modality considered, to complement our concept, we propose an initial implementation of the rendering system for a specialized vibrotactile system.
Finally, a user study is conducted to investigate the impact of different vibrotactile rendering methods, which differ in how vibrotactile information is propagated through a 3D scene.

\section{Related Works}


Haptic rendering refers to the process of simulating touch sensations in virtual environments. It broadly encompasses two modalities: kinesthetic feedback, which involves the perception of force and position through muscles and joints (e.g., force-feedback devices), and tactile feedback, which involves the stimulation of the skin through vibrations, pressure, or texture cues. Salisbury et al. (2004)~\cite{HapticRenderingDef} defined haptic rendering as ``the process by which desired sensory stimuli are imposed on the user to convey information about a virtual haptic object''” emphasizing the role of touch in perception and manipulation. Later, Laycock et al. (2007)~\cite{renderingSurvey} focused on kinesthetic rendering, describing it as ``the process of calculating a reaction force for a given position of the haptic feedback device,'' and surveyed techniques primarily involving single-point and surface rendering for devices like the Phantom series~\cite{phantom}. However, with the proliferation of smartphones, gaming consoles, and VR systems in the 2010s, tactile feedback, particularly vibrotactile rendering, has become the dominant modality. This shift has spurred research into tactile rendering techniques based on event-driven interactions and physical modeling~\cite{VibroRendering}. In this work, we focus exclusively on tactile haptic rendering, exploring how tactile signals can be generated to convey meaningful information through touch.

Tactile signals in haptic systems are typically generated using two main approaches: procedural and authored. Procedural methods rely on the simulation of physical interactions to dynamically generate haptic feedback. These are often used in contact-based scenarios where realism and precision are critical. For instance, haptic gloves and force-feedback devices simulate forces and textures based on physical models to support applications like surgical training or teleoperation~\cite{electro,reviewGloves,trainingChir,teleoperation}.
In contrast, authored approaches use pre-defined haptic effects that are manually designed and triggered based on specific events~\cite{roy2024towards} or to be synchronized with auditory feedback~\cite{Yun2023}. This methods is prevalent in consumer applications such as mobile apps, video games, and VR experiences, where vibrotactile feedback is delivered through devices like smartphones, gamepads, haptic suits, or seats~\cite{headsetKrakenRazer,owo,PneumaticSuit,freyja}. These effects are often metaphorical—representing collisions, environmental cues, or notifications—and are not generated through physical simulation.

The spatialization of haptic feedback distributing tactile signals across multiple actuators on the body has become a central topic in recent research~\cite{Danieau2018}. Multi-actuator systems are increasingly used in applications such as assistive technologies for individuals with disabilities~\cite{blind,child} and in enhancing perceptual capabilities~\cite{AuditionTrouble}. Adilkhanov et al.~\cite{surveyDevice} provide a comprehensive survey of such devices. However, spatialization introduces new challenges. Unlike visual or auditory media, haptic feedback lacks standardized models for spatial rendering. Current systems often assign effects statically to body parts without considering physical propagation or perceptual coherence over the 3D environment. This contrasts with established techniques in other domains, such as ray tracing in graphics~\cite{bookRayTracing} or spatial audio rendering~\cite{SpatialAudio}. As a result, haptic spatialization remains underdeveloped, limiting the realism and immersion of tactile experiences.

Creating compelling haptic experiences remains a complex task, particularly for non-experts. Most haptic SDKs~\cite{SkineticSDK,owoSDK,Interhaptics} are tightly coupled to specific hardware and offer limited interoperability, making cross-platform development difficult. These tools typically rely on manual authoring, where designers define and trigger haptic effects through code~\cite{roy2024towards}.
The WYVRN framework \cite{Interhaptics} exemplifies a more structured approach, drawing inspiration from audio systems in game engines. It introduces the concepts of Sources (emitters) and Body Parts (receivers), allowing developers to assign predefined effects to specific locations on the body. While it supports basic modulation, such as adjusting the intensity of the haptic feedback, it lacks support for more expressive parameters like frequency or waveform, and still requires programming expertise to implement more complex behavior.
%

Prior work has largely overlooked the role of spatial layout and object properties in shaping haptic signal propagation. To our knowledge, no existing system computes haptic feedback in 3D environments by explicitly modeling these factors. Our work addresses this gap by introducing a propagation model that accounts for the physical configuration of the scene, enabling more plausible and context-aware haptic experiences.

\section{Haptic Tracing}
\label{sec:arch}

This section introduces a novel method for rendering spatial haptic data, which enables the propagation of haptic signals through objects in the environment based on their material properties. This approach is analogous to raycasting, where light is propagated in a virtual environment according to the material properties of objects in the scene. However, unlike raycasting, which primarily propagates light through the air and not through opaque materials, haptic data is predominantly transmitted through the objects themselves. Vibrotactile signals, for instance, do not travel through the air but are conveyed through objects based on their physical characteristics, such as the density, the hardness, the elasticity or even the geometry. Additionally, while raytracing techniques are typically employed to render data for a single frame at a specific moment in time, our method aims to propagate temporal haptic signals, encompassing haptic data over a duration. This concept is akin to spatial audio rendering, which involves the propagation of audio signals within a spatial environment. Yet, audio data is mostly transmitted through the air. By leveraging the principles of both raytracing and spatial audio approaches, our work establishes a new paradigm for spatial haptic rendering, offering a comprehensive solution for immersive haptic experiences.

\begin{figure*}[h!]
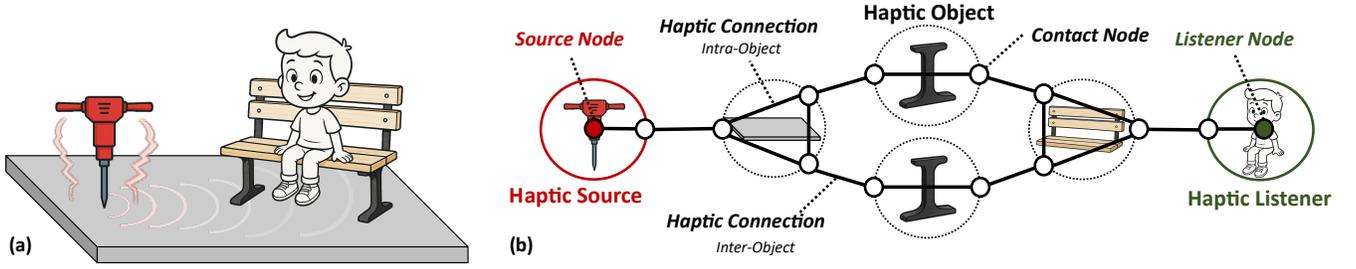

    \centering
    \begin{subfigure}[t]{.35\linewidth}
        \centering
         \includesvg[inkscapelatex=false,width=\textwidth]{Images/Scene.svg}
    \end{subfigure}
    \hfill
    \begin{subfigure}[t]{.63\linewidth}
        \centering
         \includesvg[inkscapelatex=false,width=\textwidth]{Images/HapticGraph.svg}
    \end{subfigure}
    \caption{(a) Virtual environment where a jackhammer transmits haptic information through the ground to an avatar sitting on a bench. (b) Corresponding Haptic Graph. The jackhammer (haptic source), is connected to an avatar (haptic listener) through the ground, the feet of the bench and the bench seat (haptic objects). The graph contains a source node associated to the jackhammer, a listener node associated to the avatar, and haptic nodes modeling interaction points between haptic objects. Haptic nodes are linked together through intra-connections internally within an object and through inter-connections externally.
    }
    \label{fig:haptic_graph}
\end{figure*}

\subsection{Haptic Sources, Listeners, and Objects}

Both spatial audio and raycasting share the fundamental concepts of emitters and receivers. In raycasting, light sources act as emitters, while the virtual camera serves as the recipient or receiver. Similarly, spatial audio systems rely on audio sources and audio listeners. The audio signal is modulated based on the position of the audio source, the geometry of the scene and the listener's location. Extending these principles, we introduce the notions of haptic sources and haptic listeners for spatial haptic rendering.

First, we define \textbf{Haptic Objects} as elements from the scene capable of transmitting haptic information. These elements may be bounded virtual objects or abstract volumes -- such as air or water. Haptic objects are characterized through haptic properties that govern the object’s ability to propagate haptic data. These properties include vibration transmissibility, thermal conductivity, density, hardness, and other characteristics affecting user’s perception of the object through touch.

We define \textbf{Haptic Sources} as haptic objects capable of emitting haptic signals. In this regard, different natures of haptic sources can be considered: persistent and temporary haptic sources. \textit{Persistent sources} are continuously active and associated to a virtual object in the scene, for example, a dishwasher generating vibratory data or a fire-place emitting heat. In contrast, \textit{temporary sources} are associated with short term events happening in the scene, such as a collision between two virtual objects or a sudden explosion.  

Finally, \textbf{Haptic Listeners} are haptic objects responsible of continuously sampling haptic information at a given location in the scene. These listeners act as virtual sensors strategically positioned in the virtual environment to capture haptic information to be rendered on associated haptic devices or actuators. Typically, they may be placed on an avatar (user's virtual representation) to be associated with a users' body part (e.g., index finger) or to a haptic actuator (e.g., game controller). The information that they sample is transmitted to the haptic device for its restitution. Although haptic listeners can be reconfigured at runtime, a fixed configuration (body-part / actuator) is generally enforced at the beginning of the experience to maintain consistent mapping between virtual and physical feedback. 

To illustrate these concepts, consider a user holding a virtual hammer in her left hand and striking a nail on a table. The impact generates a virtual temporary haptic source, emitting a vibration that is propagated through the hammer -- a haptic object. A haptic listener located at the user's hand captures this signal, which is then rendered on a haptic device (e.g., a game controller), allowing the user to feel the impact of the hammer on the nail.

\subsection{The Haptic Graph}
\label{sec:graph}

Given a well-defined scene, the main problem that light and audio rendering methods solve is how to efficiently compute the propagation of information (or energy) from the sources to the listeners. 
This problem requires detailed knowledge of the topology, the materials and even the dynamics of the scene, and may for instance be efficiently accomplished with raycasting techniques, propagating information through the air.
Compared to light and sound, haptic information is mostly propagated through the internal structure of objects and transmitted to other objects through direct or indirect contact. In this respect, computing the propagation of haptic data in a virtual environment also requires a precise modeling of the relationships between objects within this environment (e.g., objects that are in contact).


To efficiently compute the propagation of haptic information in a virtual scene, we introduce the concept of \textbf{Haptic Graph}. This undirected graph models how haptic signals travel through a scene by defining how haptic information is transmitted between different haptic objects and how it propagates within individual objects (see Figure~\ref{fig:haptic_graph}). The Haptic Graph is composed of haptic nodes and haptic connections (edges). A haptic node serves as the interface through which haptic data enters or exits a haptic object. A haptic node may be a source node (emitting data), a listener node (capturing data), or a contact node modeling interactions with other elements of the environment. A single haptic object can have multiple nodes, each corresponding to a distinct contact point (or area), source, or listener. Within an object, all haptic nodes are interconnected, modeling the internal pathways that allow haptic data to propagate through the object’s structure. Haptic connections define how data flows between nodes and are categorized as either inter-connections (linking nodes from different objects) or intra-connections (linking nodes within the same object). This structure allows the Haptic Graph to capture both external transmission and internal distribution of haptic signals across the scene.


The characteristics of a haptic graph can be formalized as follows. Let \( G = (V, C) \) represent the Haptic Graph of a scene with a set \(\mathcal{O}\) of haptic objects,
\begin{itemize}
\item \( V \) is the set of all haptic nodes \(v_i\) in \(G\). Haptic nodes are always associated with a single haptic object of the scene.
\item \(\mathcal{O} = \{ o_1, o_2, \dots, o_n \}\) is the set of haptic objects. Elements \(o_i\) are defined as non overlapping subsets of V such that \(\forall i \neq j, \; o_i \cap o_j = \emptyset\) and \(\bigcup_{i=1}^{n} o_i = V\).
\item \(C={\{P,T\}}\) is the set of all haptic connections \(c_{i,j}\) in the graph.
\item \( P \) is the set of intra-connections \(p_{ij}\), modeling the internal Propagation of haptic information between nodes \( v_i \) and \( v_j \) of a same haptic object. All haptic nodes associated with a haptic object are fully connected.
\(
\forall v_i, v_j \in V : (\exists p_{i,j} \in P \quad \Leftrightarrow \quad \exists o_k \in \mathcal{O} \; \text{such that} \; v_i \in o_k \; \text{and} \; v_j \in o_k)
\).
\item \( T \) is the set of inter-connections \( t_{ij} \), modeling the Transmission of haptic information between two nodes \( v_i \) and \( v_j \) belonging to two distinct haptic objects.
\(
\forall v_i, v_j \in V \; \text{with} \; v_i \in o_k \; \text{and} \; v_j  \in o_l (\exists t_{ij} \in T \quad \Rightarrow o_k \neq o_l)
\)

\end{itemize}

The Haptic Graph could contain multiple disconnected components, each representing distinct sets of haptic objects that are not interconnected. As the spatial arrangement and connectivity of objects in the virtual environment may vary over time, the Haptic Graph is dynamic and has to be updated. This ensures that changes in object positions, orientations, and contact dynamics are accurately reflected.

Figure~\ref{fig:haptic_graph} illustrates how a virtual environment (a) is modeled with a Haptic Graph (b). In this scene, a vibrating jackhammer emits haptic signals that propagate through the ground to an avatar seated on a bench. The jackhammer’s source node connects to a contact node, which transmits the signal to the ground via an inter-connection. Within the ground, haptic data travels between contact nodes through intra-connections. The ground is linked to the bench feet at two points, which in turn connect to the bench seat. The haptic signal is captured  through the listener node of the avatar, connected to the bench seat. Overall, haptic information flows through the jackhammer, ground, bench, and avatar via intra-connections, and is transferred between them using inter-connections.

\subsection{Propagation of Spatial Haptic Data}


As detailed in section \ref{sec:graph}, the propagation of haptic signals relies on a detailed understanding of the physical and topological relationships between objects in the scene. These relationships are encoded in the Haptic Graph, which models all potential pathways for haptic signals. 
The propagation of haptic information in the scene is achieved by first, finding a path or set of path connecting source nodes and listener nodes in the graph, and then, for each of these paths, computing the modulation (attenuation, degradation, amplification,etc.) of the haptic signal based on the propagation and transmission functions applied at the haptic connections.

For each pair of connected (directly or indirectly) source node and listener node, the propagation path -- or a set of relevant paths -- may be computed using path planning techniques relying on the propagation and transmission functions as heuristics. To manage computational complexity, similar to ray tracing methods that limits the number of bounces of light rays, path planning solutions may limit the number of traversed connections or nodes and stops propagation if the signal degradation reaches a given threshold. 

Now, given a path between a source and a listener, the propagation of the haptic signal can be expressed as follows. Let \( G = (V, C) \) be the Haptic Graph of the scene, where \( C = \{P, T\} \) as detailed in \ref{sec:graph}  and:

\begin{itemize}
    \item \( s \): the source node emitting the initial haptic signal.
    \item \( l \): the listener node receiving the propagated signal.
    \item \( \mathcal{P}_{s \rightarrow l} = \{v_0, v_1, \dots, v_n\} \): an ordered path of nodes from source \( s = v_0 \) to listener \( l = v_n \).
    \item \( S_0 \): the initial haptic signal emitted at node \( s \).
    \item \( f_{p_{i,j}}(S) \): the propagation function for intra-object connection \( p_{i,j} \), depending on material properties and geometry.
    \item \( f_{t_{i,j}}(S) \): the transmission function for inter-object connection \( t_{i,j} \), depending on contact parameters (e.g., contact area, pressure, orientation).
\end{itemize}

The modulation function for any haptic connection between a node \(v_i\) and a node \(v_j\) is defined as:
\[
f_{c_{i,j}}(S) =
\begin{cases}
f_{t_{i,j}}(S), & \text{if } c_{ij} \in T \quad \text{(inter-connection)} \\
f_{p_{i,j}}(S), & \text{if } c_{ij} \in P \quad \text{(intra-connection)}
\end{cases}
\]

The haptic signal received at the listener node \( l \) is computed by sequentially applying the modulation functions associated with each connection along the path \( \mathcal{P}_{s \rightarrow l} \). The final signal \( S_l \) received at the listener is given by:

\[
S_l = f_{c_{n,n-1}}(f_{c_{n-1,n-2}}(\dots f_{c_{2,1}}(f_{c_{1,0}}(S_0))\dots))
\]

This formulation captures the cumulative effect of all transformations applied to the signal as it propagates through the Haptic Graph. Function \( f_{p_{i,j}} \) and \( f_{t_{i,j}} \) may model a variety of complex modulation behaviors such as attenuation, frequency filtering, or delay for instance. 
For a given haptic modality, these functions are the same throughout the graph but are parameterized by the haptic objects' properties. However, as the propagation behavior differs from one modality to another (temperature does not propagate like vibration), dedicated functions may be defined for each modality. 

Finally, the signals received by the haptic listener from different paths (either from the same or from different sources) can be mixed by the listener before rendering. The mixing process may take various forms such as cumulative or weighted sum. Further details on the haptic mixing process are provided in Section \ref{sec:mixing}, including illustrative examples and a discussion of the method ultimately adopted.

\section{Case Study: Vibrotactile Rendering Engine}
\label{sec:impl}

This section demonstrates the use of Haptic Tracing in a vibrotactile interactive experience, where haptic feedback is used to enhance the realism and responsiveness of interactions with the scene. We detail our implementation of haptic tracing in this context and discuss a number of optimizations related to the Haptic Graph. An overview of our implementation is provided in Algorithm~\ref{alg:one}.

%
%



\begin{algorithm}[h!]
\SetAlgoNoLine
\Repeat{}{
	collisionDetection()\;
    graphUpdate()\;
	
    \For{ each Source Node $s$ in the Haptic Graph
        }{
         propagateHapticInformation(s)\; 
        }
    \For{ each Haptic Listener $l$ in the Haptic Graph
        }{
        mixHapticSignals($l$) \;
        renderHaptic($l$)\;
        }
      }      
\caption{Pseudo-code implementation of Haptic Tracing}
\label{alg:one}
\end{algorithm}




 
       


\subsection{Haptic Objects, Sources and Listeners}

Any element of the virtual environment that can transmit haptic information is considered a \textbf{Haptic Object}, in our implementation this includes virtual objects, the user representation and the air. 
\textbf{Haptic Sources} are either permanent when associated to virtual objects (e.g., jackhammer) or temporary when created following punctual events in the scene (e.g., a hammer hitting a nail). Temporary haptic sources are generated in real-time when two haptic objects collide with an energy greater than 5 joules -- equivalent to light contact. The haptic signal of a temporary source is not procedurally generated but derived from a pre-designed signal crafted by a haptic designer to simulate certain types of collisions. The pre-designed template signal is modulated based on the physical characteristics of the collision as expressed in Equation~\ref{eq:templateModulation}. 
%

\begin{equation}
\label{eq:templateModulation}
S_s = S_t\times\frac{\frac{1}{2} m_1 v_1^2 + \frac{1}{2} m_2 v_2^2}{E_{max}}
\end{equation}
where \(S_s\) is the resulting signal of the source, \(S_t\) is the template signal, \(m_1\) and \(m_2\) denote the masses of the two interacting objects, \(v_1\) and \(v_2\) represent their respective velocities, and \(E_{max}\) the maximum deliverable energy. In our implementation, \(E_{max}\) is fixed at 100 joules.

%
Finally, the number and positions of \textbf{Haptic Listeners} are defined based on the available body parts of the user's avatar. In our implementation, we define 25 default body parts of varying sizes. A default configuration employs 10 listener locations (see Section~\ref{sec:eval}), distributed across different regions of the user’s body. Each listener is associated with a bounding volume that delimits its spatial extent.
\subsection{Haptic Graph}

As defined in Section~\ref{sec:graph}, the \textbf{Haptic Graph} is composed of a set of haptic nodes and a set of haptic connections \( G = (V, C) \).
In our implementation, the $graphUpdate()$  process in Algorithm~\ref{alg:one} updates the Haptic Graph to reflect changes in the environment detected by the $collisionDetection()$ process. When two haptic objects \( o_i\) and  \(o_j \) come into contact, two contact nodes \( v_k\) and \(v_l \) are added to the graph, along with inter-connection \( c_{k,l} \) and the intra-connections with the other nodes in \( o_i\) and  \(o_j \). When the two objects are no longer in contact, these haptic nodes and connections are removed from the graph. 
%
The relationships between haptic objects are tracked using the physics engine of the 3D environment (Unity 3D in our case), which detects when objects come into contact.

\subsection{Propagation of Spatial Haptic Data}
After updating the Haptic Graph, haptic data is propagated through its nodes (propagateHapticInformation(s) in Algorithm~\ref{alg:one}). 
This step is a crucial component of haptic tracing as it significantly affects both rendering quality and performance. 
We propose two distinct propagation strategies. The first, Breadth-First Propagation, especially relevant for persistent haptic sources, is inspired by the Breadth-First Search algorithm. 
It iteratively transmits Information emitted by each source to neighboring nodes until propagation naturally terminates. 
This exhaustive approach allows to identifies all possible paths from a source to any reachable listener.
The second strategy, better suited for complex environments or transient sources, computes optimal paths individually for each source-listener pair. 
Based on Dijkstra’s algorithm, this dynamic programming method limits propagation to paths that minimize signal attenuation, focusing computational effort on the most efficient routes. The propagation and transmission functions serve as cost heuristics to guide this path planning process.

Propagation and transmission functions are vital components of haptic tracing. Inspired by the concepts of shaders in computer graphics, our implementation allows designers to select from predefined functions or implement custom ones.
For intra-connections,
we provide a basic set of distance-based attenuation functions (e.g., constant, linear, exponential), as well as a more advanced propagation function (see Equation~\ref{fct1}) that accounts for both the distance between the contact nodes and object-specific haptic properties for finer control.



\begin{equation}
\label{fct1}
f_{p_{i,j}}(x)= x.e^{- \max((1-\kappa)\rho,\epsilon)d}.(1-\kappa).\rho
\end{equation}

Where \(x\) is a sample of the input signal, \(d\) represents the distance between the two contact nodes in the 3D space, \(\kappa\) the vibrotactile transmissability coefficient of the haptic object, \(\rho\) the density coefficient, and \(\epsilon\) is a minimal threshold for the propagation -- with a $0.1$ default value.

For inter-object connections, the system allows the designer to apply any existing or custom function, depending on the specific needs and the nature of the expected interaction. By default, no modulation is applied to this type of connection, under the assumption that, in our implementation, the most significant attenuation of haptic information occurs during intra-object transmission.

\subsection{Mixing Spatial Haptic Data}
\label{sec:mixing}

Listener nodes may capture haptic information coming from different paths, either from the same source or from different ones. This overlap can overwhelm the user's sensory system, potentially resulting in an unpleasant, confusing, or even disturbing experience. To mitigate this, the haptic listener can mix incoming signals before rendering ($mixHapticSignals(l)$ process in Algorithm~\ref{alg:one}).

Our system implements a default method based on cumulative aggregation and normalization. As described in Equation \ref{eq:cumulative}, it modulates the contribution of each haptic source in a balanced manner.

\begin{equation}
\label{eq:cumulative}
s_{\text{final}} =
\begin{cases}
s_{\text{sum}}, & \text{if } s_{\text{sum}} \leq s_{\text{max}} \\
\frac{s_{\text{sum}}}{\|s_{\text{sum}}\|} \cdot s_{\text{max}}, & \text{if } s_{\text{sum}} > s_{\text{max}}
\end{cases}
\end{equation}
 
\begin{itemize}
\item \( S = \{s_1, s_2, \dots, s_n\} \) the set of individual haptic signals.  
\item \( s_{\text{sum}} = \sum_{i=1}^{n} s_i \) the total combined signal,  
\item  \( s_{\text{max}} \) the maximum allowable haptic intensity.  
\item \( s_{\text{final}} \) the resulting signal after normalization.
\end{itemize}


%

Alternative custom mixing functions may be provided by the designer to accommodate the characteristics and constraints of specific uses cases or different haptic modalities. Other mixing strategies may use priority coefficients to weight the importance of the different sources, or include decay to gradually reduce the impact of persistent sources. 






\section{Evaluation}
\label{sec:eval}

The experiment aims to evaluate how the user perceives different spatialized haptic rendering methods (vibrotactile feedback) from a multisensory integration perspective (i.e., evaluate the coherence between visual and haptic feedback). Precisely, we investigate the impact of different vibrotactile rendering methods, which differ in how vibrotactile information is propagated through a virtual 3D environment.
The evaluation focuses on how the information is propagated from the haptic source to the haptic listener, the number of actuators that are triggered and the attenuation function (see Section~\ref{sec:modalities}).
Our primary hypothesis is that spatialized methods leveraging scene information will improve the users' haptic experience. 






\subsection{Apparatus and Participants}

We used commercial-available vibrotactile devices in order to deliver spatialized haptic rendering. Precisely, we used 3 different devices: (1) the Razer Kraken V4 Pro headset, with one vibrotactile actuator per ear, (2) the Razer Wolverine V3 Pro gamepad with two vibrotactile actuators on each side (left and right hand), and (3) the Razer Freyja cushion placed on an office chair with four actuators on the back and two on the bottom. 
One haptic listener was associated to each actuator (10 in total), which enabled the individual control of each one.

41 participants were included in the study, ($M=29, SD=9.43$) including three women. no specific inclusion criteria was considered. The experiment received the approval of the ethical committee of [anonymized].


%

%





\begin{figure}[b]
  \includegraphics[width=0.95\linewidth]{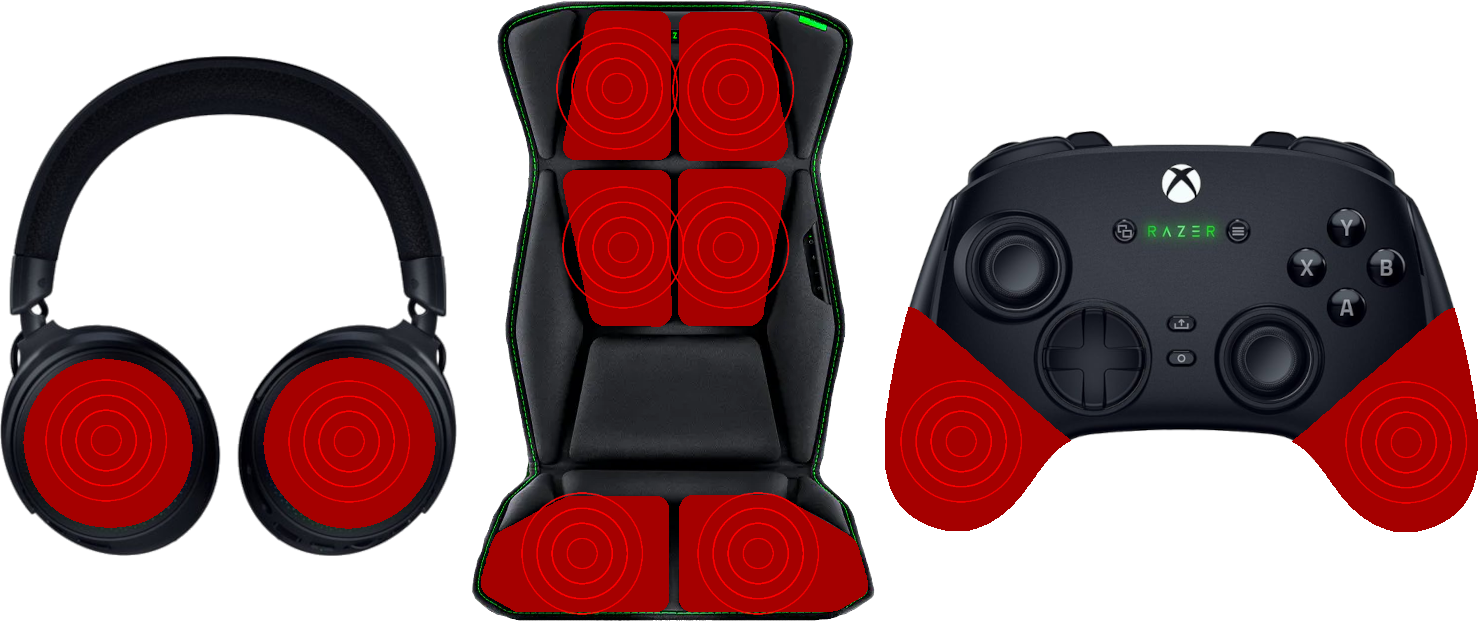}
  \caption{Experimental Setup composed of three independent devices, the red area denote the placements of the actuators of each devices. From left to right, Razer Kraken V4 Pro headset (head, left and right); Razer Freyja cushion (upper back, middle back and upper leg, left and right); Wolverine V3 Pro gamepad (hand, left and right).}
  \label{Device}
\end{figure}

\subsection{Experimental Design}

\subsubsection{Scenario}
\label{sec:scenario}

To ensure participants were exposed to the same haptic effects, we considered a passive scenario in which the user is transported through a mine aboard a cart (see accompanying video). Throughout the travel, participants encounter a series of events that generated haptic feedback. The scenario is structured into the following six sections:

\begin{enumerate}
    \item Stalactites fall to the ground and generate a haptic source at the collision. The scenario contains 15 stalactites. Stalactite has different sizes and weights.
    \item The cart traverses a bridge and an upward wind stream comes from the bottom of the cliff. The intensity of the wind stream varies during the traversal.
    \item Two elves shoot the user using blowguns. When the projectile and the user collide, a haptic source is generated. The elves shoot multiple bullets every five seconds.
    \item An explosion occurs. The shock wave traverses the user from left to right. 
    \item Two mine carts cross before and behind the user. The vibration of the wheels on the rails is transmitted to the user.
    \item The user goes through nine walls with different haptic properties: three stone walls, three wood walls, and three ribbon walls. At the collision between the user and the wall, a haptic source is generated.
\end{enumerate}



\subsubsection{Rendering Modalities}
\label{sec:modalities}

Four rendering modalities were defined altering the propagation and modulation of the haptic information. 
\paragraph{Haptic Tracing (HT)} The first modality tested is the complete version of the haptic tracing implementation presented in Section~\ref{sec:impl}. All functionalities of the haptic graph are used and the customization of haptic information is performed with custom filters using haptic properties of objects. 
%
%
\paragraph{Single actuator, distance attenuation (SD)} Only the nearest haptic listener to the haptic source will be enabled (i.e., the haptic graph is only updated considering the haptic listener closer to the current haptic source). Moreover, haptic information attenuation only considers the distance (and not the haptic properties of the objects). The attenuation function used (see Function~\ref{fct2}) is a simplified version of Function~\ref{fct1} in which only the distance is used to modulate the haptic information ($\alpha=0.2$).

\begin{equation}
\label{fct2}
f_{p_{i,j}}(x) = x.e^{- \alpha d}
\end{equation}

%
%
%
\paragraph{Multiple actuators, distance attenuation (MD)} This modality renders haptic information on all haptic listeners and haptic signals are attenuated using a distance filter defined by Function~\ref{fct2}. For this modality, a different graph connectivity is defined, in which all haptic sources are connected to all haptic listeners. 




\paragraph{Multiple actuators, no attenuation (MN)} This version behaves as the previous condition, but no attenuation is applied. This rendering method will just render the haptic signal in all haptic devices without any modulation. 
This modality represents the simplest method for creating a haptic experience in which no prior information can be drawn from the layout of the 3D scene. 

\subsection{Experimental Protocol}

Upon arrival, participants read and signed the consent form, which explained the purpose of the experiment and their rights. The experiment was divided into four different blocks, one for each rendering modality. The order of the conditions was randomized. 
For each condition, the experimenter ensured that all the haptic devices were properly set up and that the participant could perceive the haptic feedback. Then, the automatic exploration of the mine environment started, which lasted 2 minutes and 30 seconds. Once finished, participants had to complete the haptic experience questionnaire~\cite{HXQuestionnaire}. The questionnaire evaluates the quality and interest of haptic experiences according to four criteria: realism, harmony, involvement, and expressivity. Participants were also informally interviewed at the end of the experiment for additional feedback and asked to rank the four conditions by preference.

\subsection{Results}


\begin{figure}[t]
    \centering
    \includegraphics[width=1\linewidth]{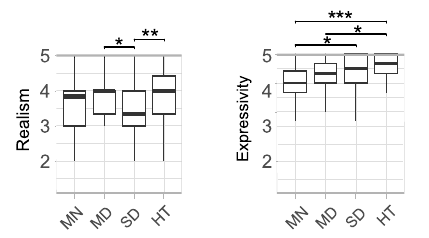}
    \caption{Boxplots for the HX questionnaire results for realism and expressivity (* p<0.05, *** p<0.001).}
    \label{fig:post_hocs}
\end{figure}

One participant was removed from the analysis as she/he only selected the maximum values for every Likert-scales. Thus, the total number of participants included in the analysis was 40.

First, regarding the \textbf{preferred condition}, 19 participants selected the HT condition as the preferred one, followed by MD with 10, the MN with 8 and the SD with 4. 
While the HT condition was the most preferred, chosen by nearly half of the participants, the two other conditions leveraging multiple actuators were selected by a total of 44\% of participants. 
This result tends to support that HT is preferred over the other tested methods, but methods using multiple actuators were also well appreciated.

%
Then, answers from the \textbf{HX questionnaire} were aggregated for each dimension: realism, harmony, involvement, and expressivity. Non-parametric Friedman tests were conducted and Wilcoxon signed-rank test with Bonferroni correction were used for post-hoc analysis. 
The Friedman analysis only showed significant results for realism ({\small $\chi^2(3,40)=15.5;p<0.01$}) and expressivity ({\small $\chi^2(3,40)=14.1;p<0.01$}).
Post-hoc tests showed that the SD condition was ranked less realistic than MD and HT conditions, while for expressivity, HT was significantly higher than MN and MD conditions, and SD was also significantly higher than MN (see Figure~\ref{fig:post_hocs}). 
These results suggest that the SD condition was considered the least realistic, maybe due to the focalization of the feedback in a single actuator, and that spatial and intensity modulation effectively generates more variability in the perceived sensations.

\subsection{Dicussion}

The aim of the experiment was to explore the impact of spatialization and modulation on the overall appreciation of the haptic experience. The results show that leveraging multiple actuators is indeed beneficial for overall appreciation (in terms of preference and realism). However, using the scene layout (HT) tends to further increase the level of appreciation. These results suggest that users appreciated the spatialization of the haptic feedback, which generated an overall more believable experience. However, the use of multiple actuators can be detrimental to the expressivity of the haptic feedback if haptic signals are not modulated or properly spatialized. This can be seen in the expressivity scores, where HT and SD achieved higher scores. Regarding the other dimensions of the HX questionnaire (harmony and involvement), the fact that no significant differences were observed can be linked to the visuo-haptic synchronization and the fact that the user was passive in the environment. In this respect, future work could explore how interaction could modulate the obtained results.

During the experiment, we also observed several interesting individual preferences. Some users were occasionally unsettled due to the propagation of effects in the HT condition. For example, in the case of a stalactite falling to the ground, the realistic vibration propagation in HT caused the vibration to transfer through the ground, stimulating both legs of the user. Some participants preferred a more simplified propagation, where a falling stalactite on the right would only stimulate the right leg. While this more localized response is not realistic from a physical perspective, it may align better with users' intuitive expectations. Nevertheless, these observations confirm that users tend to favor plausible effects over realistic ones. Other users preferred more intense haptic feedback, thus favoring the MD and MN versions. However, it remains unclear how this preference would last during a long-lasting experience, in which high-intensity vibrations could become tiresome or even disruptive. Both examples illustrate the need to further assess user haptic experience during prolonged sessions and highlight the need to consider user preferences when crafting haptic experiences.

The presented experiment also has a number of limitations that will require further work. First, while we adopted specific propagation and mixing approaches, it is important to note that alternative strategies could be defined and empirically evaluated. In our experimental setup, only one haptic effect was presented at a time, which limited the need for mixing the haptic information. However, in richer haptic experiences involving concurrent stimuli, the choice of mixing strategy may significantly impact user perception and comfort. Moreover, MD and MN will most probably underperform in such scenarios due to the difficulty for the user to match visual and haptic stimuli. Future work could explore and assess different mixing techniques to identify optimal solutions for more immersive scenarios. Second, we considered a fixed hardware setup; different hardware setups involving additional body parts or combining multiple tactile modalities should be further explored. Finally, we did not explore the added value of the proposed haptic rendering method from the authoring perspective. Further work should explore how haptic designers could leverage Haptic Tracing to create more engaging and immersive experiences.

\section{Conclusion}

In this paper, we introduced a novel method for spatial haptic rendering that simplifies the creation of interactive haptic experiences without relying on physical simulations. By leveraging concepts from ray tracing and spatial audio, our proposed system enables dynamic and perceptually rich interactions in 3D environments. Through the integration of vibrotactile feedback in a fully functional system, we demonstrate how our framework can deliver more responsive and spatially coherent haptic experiences. The results of our user study indicate that Haptic Tracing significantly improves users' haptic experience, highlighting its potential to bridge the gap between artistic expressivity and realistic haptic feedback. This work paves the way for more accessible and versatile haptic systems, fostering innovative applications in interactive virtual experiences.

\bibliographystyle{ACM-Reference-Format}
\bibliography{main}


\begin{thebibliography}{35}


\ifx \showCODEN    \undefined \def \showCODEN     #1{\unskip}     \fi
\ifx \showDOI      \undefined \def \showDOI       #1{#1}\fi
\ifx \showISBNx    \undefined \def \showISBNx     #1{\unskip}     \fi
\ifx \showISBNxiii \undefined \def \showISBNxiii  #1{\unskip}     \fi
\ifx \showISSN     \undefined \def \showISSN      #1{\unskip}     \fi
\ifx \showLCCN     \undefined \def \showLCCN      #1{\unskip}     \fi
\ifx \shownote     \undefined \def \shownote      #1{#1}          \fi
\ifx \showarticletitle \undefined \def \showarticletitle #1{#1}   \fi
\ifx \showURL      \undefined \def \showURL       {\relax}        \fi
\providecommand\bibfield[2]{#2}
\providecommand\bibinfo[2]{#2}
\providecommand\natexlab[1]{#1}
\providecommand\showeprint[2][]{arXiv:#2}

\bibitem[Actronika(2025)]%
        {SkineticSDK}
\bibfield{author}{\bibinfo{person}{Actronika}.}
  \bibinfo{year}{2025}\natexlab{}.
\newblock \bibinfo{title}{Unitouch SDK}.
\newblock
\newblock
\urldef\tempurl%
\url{https://unitouch.actronika.com/SDK/intro/}
\showURL{%
\tempurl}
\newblock
\shownote{Consulté le 12 mai 2025}.


\bibitem[Adilkhanov et~al\mbox{.}(2022)]%
        {surveyDevice}
\bibfield{author}{\bibinfo{person}{Adilzhan Adilkhanov},
  \bibinfo{person}{Matteo Rubagotti}, {and} \bibinfo{person}{Zhanat
  Kappassov}.} \bibinfo{year}{2022}\natexlab{}.
\newblock \showarticletitle{Haptic Devices: Wearability-Based Taxonomy and
  Literature Review}.
\newblock \bibinfo{journal}{\emph{IEEE Access}}  \bibinfo{volume}{10}
  (\bibinfo{year}{2022}), \bibinfo{pages}{91923--91947}.
\newblock
\urldef\tempurl%
\url{https://doi.org/10.1109/ACCESS.2022.3202986}
\showDOI{\tempurl}


\bibitem[Altinsoy and Merchel(2012)]%
        {electro}
\bibfield{author}{\bibinfo{person}{M.~Ercan Altinsoy} {and}
  \bibinfo{person}{Sebastian Merchel}.} \bibinfo{year}{2012}\natexlab{}.
\newblock \showarticletitle{Electrotactile Feedback for Handheld Devices with
  Touch Screen and Simulation of Roughness}.
\newblock \bibinfo{journal}{\emph{IEEE Transactions on Haptics}}
  \bibinfo{volume}{5}, \bibinfo{number}{1} (\bibinfo{year}{2012}),
  \bibinfo{pages}{6--13}.
\newblock
\urldef\tempurl%
\url{https://doi.org/10.1109/TOH.2011.56}
\showDOI{\tempurl}


\bibitem[and(2021)]%
        {AuditionTrouble}
\bibfield{author}{\bibinfo{person}{Mark D.~Fletcher and}.}
  \bibinfo{year}{2021}\natexlab{}.
\newblock \showarticletitle{Using haptic stimulation to enhance auditory
  perception in hearing-impaired listeners}.
\newblock \bibinfo{journal}{\emph{Expert Review of Medical Devices}}
  \bibinfo{volume}{18}, \bibinfo{number}{1} (\bibinfo{year}{2021}),
  \bibinfo{pages}{63--74}.
\newblock
\urldef\tempurl%
\url{https://doi.org/10.1080/17434440.2021.1863782}
\showDOI{\tempurl}
\showeprint{https://doi.org/10.1080/17434440.2021.1863782}
\newblock
\shownote{PMID: 33372550}.


\bibitem[Anwar et~al\mbox{.}(2023)]%
        {HXQuestionnaire}
\bibfield{author}{\bibinfo{person}{Ahmed Anwar}, \bibinfo{person}{Tianzheng
  Shi}, {and} \bibinfo{person}{Oliver Schneider}.}
  \bibinfo{year}{2023}\natexlab{}.
\newblock \showarticletitle{Factors of Haptic Experience across Multiple Haptic
  Modalities}. In \bibinfo{booktitle}{\emph{Proceedings of the 2023 CHI
  Conference on Human Factors in Computing Systems}} (Hamburg, Germany)
  \emph{(\bibinfo{series}{CHI '23})}. \bibinfo{publisher}{Association for
  Computing Machinery}, \bibinfo{address}{New York, NY, USA}, Article
  \bibinfo{articleno}{260}, \bibinfo{numpages}{12}~pages.
\newblock
\showISBNx{9781450394215}
\urldef\tempurl%
\url{https://doi.org/10.1145/3544548.3581514}
\showDOI{\tempurl}


\bibitem[Brahimaj et~al\mbox{.}(2024)]%
        {headsetKrakenRazer}
\bibfield{author}{\bibinfo{person}{Detjon Brahimaj}, \bibinfo{person}{Eric
  Vezzoli}, \bibinfo{person}{Fr{\'e}d{\'e}ric Giraud}, {and}
  \bibinfo{person}{Betty Semail}.} \bibinfo{year}{2024}\natexlab{}.
\newblock \showarticletitle{Enhancing Object Localization in VR: Tactile-Based
  HRTF and Vibration Headphones for Spatial Haptic Feedback}.
\newblock \bibinfo{journal}{\emph{IEEE Transactions on Haptics (ToH)}}
  (\bibinfo{year}{2024}).
\newblock


\bibitem[Cirio et~al\mbox{.}(2013)]%
        {VibroRendering}
\bibfield{author}{\bibinfo{person}{Gabriel Cirio}, \bibinfo{person}{Maud
  Marchal}, \bibinfo{person}{Anatole Lécuyer}, {and}
  \bibinfo{person}{Jeremy~R. Cooperstock}.} \bibinfo{year}{2013}\natexlab{}.
\newblock \showarticletitle{Vibrotactile Rendering of Splashing Fluids}.
\newblock \bibinfo{journal}{\emph{IEEE Transactions on Haptics}}
  \bibinfo{volume}{6}, \bibinfo{number}{1} (\bibinfo{year}{2013}),
  \bibinfo{pages}{117--122}.
\newblock
\urldef\tempurl%
\url{https://doi.org/10.1109/TOH.2012.34}
\showDOI{\tempurl}


\bibitem[Culbertson et~al\mbox{.}(2018)]%
        {culbertson2018haptics}
\bibfield{author}{\bibinfo{person}{Heather Culbertson},
  \bibinfo{person}{Samuel~B Schorr}, {and} \bibinfo{person}{Allison~M
  Okamura}.} \bibinfo{year}{2018}\natexlab{}.
\newblock \showarticletitle{Haptics: The present and future of artificial touch
  sensation}.
\newblock \bibinfo{journal}{\emph{Annual review of control, robotics, and
  autonomous systems}} \bibinfo{volume}{1}, \bibinfo{number}{1}
  (\bibinfo{year}{2018}), \bibinfo{pages}{385--409}.
\newblock


\bibitem[Daeseok et~al\mbox{.}(2023)]%
        {PneumaticSuit}
\bibfield{author}{\bibinfo{person}{Kang Daeseok}, \bibinfo{person}{Chang-Gyu
  Lee}, {and} \bibinfo{person}{Ohung Kwon}.} \bibinfo{year}{2023}\natexlab{}.
\newblock \showarticletitle{Pneumatic and acoustic suit: multimodal haptic suit
  for enhanced virtual reality simulation}.
\newblock \bibinfo{journal}{\emph{Virtual Reality}}  \bibinfo{volume}{27}
  (\bibinfo{date}{02} \bibinfo{year}{2023}).
\newblock
\urldef\tempurl%
\url{https://doi.org/10.1007/s10055-023-00756-5}
\showDOI{\tempurl}


\bibitem[Danieau et~al\mbox{.}(2018)]%
        {Danieau2018}
\bibfield{author}{\bibinfo{person}{Fabien Danieau}, \bibinfo{person}{Philippe
  Guillotel}, \bibinfo{person}{Olivier Dumas}, \bibinfo{person}{Thomas Lopez},
  \bibinfo{person}{Bertrand Leroy}, {and} \bibinfo{person}{Nicolas Mollet}.}
  \bibinfo{year}{2018}\natexlab{}.
\newblock \showarticletitle{HFX studio: haptic editor for full-body immersive
  experiences}. In \bibinfo{booktitle}{\emph{Proceedings of the 24th ACM
  Symposium on Virtual Reality Software and Technology}} (Tokyo, Japan)
  \emph{(\bibinfo{series}{VRST '18})}. \bibinfo{publisher}{Association for
  Computing Machinery}, \bibinfo{address}{New York, NY, USA}, Article
  \bibinfo{articleno}{37}, \bibinfo{numpages}{9}~pages.
\newblock
\showISBNx{9781450360869}
\urldef\tempurl%
\url{https://doi.org/10.1145/3281505.3281518}
\showDOI{\tempurl}


\bibitem[Gani et~al\mbox{.}(2022)]%
        {trainingChir}
\bibfield{author}{\bibinfo{person}{Abrar Gani}, \bibinfo{person}{Oliver
  Pickering}, \bibinfo{person}{Caroline Ellis}, \bibinfo{person}{Omar Sabri},
  {and} \bibinfo{person}{Philip Pucher}.} \bibinfo{year}{2022}\natexlab{}.
\newblock \showarticletitle{Impact of haptic feedback on surgical training
  outcomes: A Randomised Controlled Trial of haptic versus non-haptic immersive
  virtual reality training}.
\newblock \bibinfo{journal}{\emph{Annals of Medicine and Surgery}}
  \bibinfo{volume}{83} (\bibinfo{year}{2022}), \bibinfo{pages}{104734}.
\newblock
\showISSN{2049-0801}
\urldef\tempurl%
\url{https://doi.org/10.1016/j.amsu.2022.104734}
\showDOI{\tempurl}


\bibitem[Glassner(1989)]%
        {bookRayTracing}
\bibfield{author}{\bibinfo{person}{Andrew~S Glassner}.}
  \bibinfo{year}{1989}\natexlab{}.
\newblock \bibinfo{booktitle}{\emph{An introduction to ray tracing}}.
\newblock \bibinfo{publisher}{Morgan Kaufmann}.
\newblock


\bibitem[Guillotel et~al\mbox{.}(2025)]%
        {guillotel2025adding}
\bibfield{author}{\bibinfo{person}{Philippe Guillotel},
  \bibinfo{person}{Yeshwant Muthusamy}, \bibinfo{person}{Quentin Galvane},
  \bibinfo{person}{Eric Vezzoli}, \bibinfo{person}{Lars Nockenberg},
  \bibinfo{person}{Iraj Sodagar}, \bibinfo{person}{Henry Da~Costa},
  \bibinfo{person}{Alexandre Hulsken}, \bibinfo{person}{Gurvan Lecuyer},
  \bibinfo{person}{Matthieu~Perreira Da~Silva}, {et~al\mbox{.}}}
  \bibinfo{year}{2025}\natexlab{}.
\newblock \showarticletitle{Adding Touch to Immersive Media: An Overview of the
  MPEG Haptics Coding Standard}.
\newblock \bibinfo{journal}{\emph{IEEE Transactions on Haptics}}
  (\bibinfo{year}{2025}).
\newblock


\bibitem[Huppert et~al\mbox{.}(2021)]%
        {blind}
\bibfield{author}{\bibinfo{person}{Felix Huppert}, \bibinfo{person}{Gerold
  Hoelzl}, {and} \bibinfo{person}{Matthias Kranz}.}
  \bibinfo{year}{2021}\natexlab{}.
\newblock \showarticletitle{GuideCopter - A Precise Drone-Based Haptic Guidance
  Interface for Blind or Visually Impaired People}. In
  \bibinfo{booktitle}{\emph{Proceedings of the 2021 CHI Conference on Human
  Factors in Computing Systems}} (Yokohama, Japan) \emph{(\bibinfo{series}{CHI
  '21})}. \bibinfo{publisher}{Association for Computing Machinery},
  \bibinfo{address}{New York, NY, USA}, Article \bibinfo{articleno}{218},
  \bibinfo{numpages}{14}~pages.
\newblock
\showISBNx{9781450380966}
\urldef\tempurl%
\url{https://doi.org/10.1145/3411764.3445676}
\showDOI{\tempurl}


\bibitem[Kim and Schneider(2020)]%
        {kim2020defining}
\bibfield{author}{\bibinfo{person}{Erin Kim} {and} \bibinfo{person}{Oliver
  Schneider}.} \bibinfo{year}{2020}\natexlab{}.
\newblock \showarticletitle{Defining haptic experience: foundations for
  understanding, communicating, and evaluating HX}. In
  \bibinfo{booktitle}{\emph{Proceedings of the 2020 CHI conference on human
  factors in computing systems}}. \bibinfo{pages}{1--13}.
\newblock


\bibitem[Laycock and Day(2007a)]%
        {renderingSurvey}
\bibfield{author}{\bibinfo{person}{Stephen Laycock} {and} \bibinfo{person}{A.
  Day}.} \bibinfo{year}{2007}\natexlab{a}.
\newblock \showarticletitle{A Survey of Haptic Rendering Techniques}.
\newblock \bibinfo{journal}{\emph{Computer Graphics Forum}}
  \bibinfo{volume}{26} (\bibinfo{date}{03} \bibinfo{year}{2007}),
  \bibinfo{pages}{50 -- 65}.
\newblock
\urldef\tempurl%
\url{https://doi.org/10.1111/j.1467-8659.2007.00945.x}
\showDOI{\tempurl}


\bibitem[Laycock and Day(2007b)]%
        {laycock2007survey}
\bibfield{author}{\bibinfo{person}{Stephen~D Laycock} {and} \bibinfo{person}{AM
  Day}.} \bibinfo{year}{2007}\natexlab{b}.
\newblock \showarticletitle{A survey of haptic rendering techniques}. In
  \bibinfo{booktitle}{\emph{Computer graphics forum}},
  Vol.~\bibinfo{volume}{26}. Wiley Online Library, \bibinfo{pages}{50--65}.
\newblock


\bibitem[Owo(2025a)]%
        {owoSDK}
\bibfield{author}{\bibinfo{person}{Owo}.} \bibinfo{year}{2025}\natexlab{a}.
\newblock \bibinfo{title}{Owo SDK}.
\newblock
\newblock
\urldef\tempurl%
\url{https://owogame.com/developers/}
\showURL{%
\tempurl}
\newblock
\shownote{Consulté le 12 mai 2025}.


\bibitem[Owo(2025b)]%
        {owo}
\bibfield{author}{\bibinfo{person}{Owo}.} \bibinfo{year}{2025}\natexlab{b}.
\newblock \bibinfo{title}{Owo suit}.
\newblock
\newblock
\urldef\tempurl%
\url{https://owogame.com/product/}
\showURL{%
\tempurl}
\newblock
\shownote{Consulted may 12th 2025}.


\bibitem[Pacchierotti et~al\mbox{.}(2017)]%
        {Pacchierotti17}
\bibfield{author}{\bibinfo{person}{Claudio Pacchierotti},
  \bibinfo{person}{Stephen Sinclair}, \bibinfo{person}{Massimiliano Solazzi},
  \bibinfo{person}{Antonio Frisoli}, \bibinfo{person}{Vincent Hayward}, {and}
  \bibinfo{person}{Domenico Prattichizzo}.} \bibinfo{year}{2017}\natexlab{}.
\newblock \showarticletitle{Wearable Haptic Systems for the Fingertip and the
  Hand: Taxonomy, Review, and Perspectives}.
\newblock \bibinfo{journal}{\emph{IEEE Transactions on Haptics}}
  \bibinfo{volume}{10}, \bibinfo{number}{4} (\bibinfo{year}{2017}),
  \bibinfo{pages}{580--600}.
\newblock
\urldef\tempurl%
\url{https://doi.org/10.1109/TOH.2017.2689006}
\showDOI{\tempurl}


\bibitem[Park et~al\mbox{.}(2021)]%
        {child}
\bibfield{author}{\bibinfo{person}{Wanjoo Park}, \bibinfo{person}{Vahan
  Babushkin}, \bibinfo{person}{Samra Tahir}, {and} \bibinfo{person}{Mohamad
  Eid}.} \bibinfo{year}{2021}\natexlab{}.
\newblock \showarticletitle{Haptic Guidance to Support Handwriting for Children
  With Cognitive and Fine Motor Delays}.
\newblock \bibinfo{journal}{\emph{IEEE Transactions on Haptics}}
  \bibinfo{volume}{14}, \bibinfo{number}{3} (\bibinfo{year}{2021}),
  \bibinfo{pages}{626--634}.
\newblock
\urldef\tempurl%
\url{https://doi.org/10.1109/TOH.2021.3068786}
\showDOI{\tempurl}


\bibitem[Perret and Vander~Poorten(2018)]%
        {reviewGloves}
\bibfield{author}{\bibinfo{person}{J. Perret} {and} \bibinfo{person}{E.
  Vander~Poorten}.} \bibinfo{year}{2018}\natexlab{}.
\newblock \showarticletitle{Touching Virtual Reality: A Review of Haptic
  Gloves}. In \bibinfo{booktitle}{\emph{ACTUATOR 2018; 16th International
  Conference on New Actuators}}. \bibinfo{pages}{1--5}.
\newblock


\bibitem[Razer(2025)]%
        {freyja}
\bibfield{author}{\bibinfo{person}{Razer}.} \bibinfo{year}{2025}\natexlab{}.
\newblock \bibinfo{title}{Freyja Cushion}.
\newblock
\newblock
\urldef\tempurl%
\url{https://www.razer.com/fr-fr/gaming-chairs-accessories/razer-freyja}
\showURL{%
\tempurl}
\newblock
\shownote{Consulted may 12th 2025}.


\bibitem[Roy et~al\mbox{.}(2024)]%
        {roy2024towards}
\bibfield{author}{\bibinfo{person}{Tom Roy}, \bibinfo{person}{Yann
  Gl{\'e}marec}, \bibinfo{person}{Gurvan L{\'e}cuyer}, \bibinfo{person}{Quentin
  Galvane}, \bibinfo{person}{Philippe Guillotel}, {and} \bibinfo{person}{Ferran
  Argelaguet}.} \bibinfo{year}{2024}\natexlab{}.
\newblock \showarticletitle{Towards End-User Customization of Haptic
  Experiences}. In \bibinfo{booktitle}{\emph{International Conference on Human
  Haptic Sensing and Touch Enabled Computer Applications}}. Springer,
  \bibinfo{pages}{411--425}.
\newblock


\bibitem[Salisbury et~al\mbox{.}(2004)]%
        {HapticRenderingDef}
\bibfield{author}{\bibinfo{person}{K. Salisbury}, \bibinfo{person}{F. Conti},
  {and} \bibinfo{person}{F. Barbagli}.} \bibinfo{year}{2004}\natexlab{}.
\newblock \showarticletitle{Haptic rendering: introductory concepts}.
\newblock \bibinfo{journal}{\emph{IEEE Computer Graphics and Applications}}
  \bibinfo{volume}{24}, \bibinfo{number}{2} (\bibinfo{year}{2004}),
  \bibinfo{pages}{24--32}.
\newblock
\urldef\tempurl%
\url{https://doi.org/10.1109/MCG.2004.1274058}
\showDOI{\tempurl}


\bibitem[Schissler and Manocha(2017)]%
        {SpatialAudio}
\bibfield{author}{\bibinfo{person}{Carl Schissler} {and}
  \bibinfo{person}{Dinesh Manocha}.} \bibinfo{year}{2017}\natexlab{}.
\newblock \showarticletitle{Interactive Sound Propagation and Rendering for
  Large Multi-Source Scenes}.
\newblock \bibinfo{journal}{\emph{ACM Trans. Graph.}} \bibinfo{volume}{36},
  \bibinfo{number}{4}, Article \bibinfo{articleno}{114c} (\bibinfo{date}{July}
  \bibinfo{year}{2017}), \bibinfo{numpages}{12}~pages.
\newblock
\showISSN{0730-0301}
\urldef\tempurl%
\url{https://doi.org/10.1145/3072959.2943779}
\showDOI{\tempurl}


\bibitem[Schneider and MacLean(2016)]%
        {Schneider2016}
\bibfield{author}{\bibinfo{person}{Oliver~S. Schneider} {and}
  \bibinfo{person}{Karon~E. MacLean}.} \bibinfo{year}{2016}\natexlab{}.
\newblock \showarticletitle{Studying design process and example use with
  Macaron, a web-based vibrotactile effect editor}. In
  \bibinfo{booktitle}{\emph{2016 IEEE Haptics Symposium (HAPTICS)}}.
  \bibinfo{pages}{52--58}.
\newblock
\urldef\tempurl%
\url{https://doi.org/10.1109/HAPTICS.2016.7463155}
\showDOI{\tempurl}


\bibitem[Singh et~al\mbox{.}(2020)]%
        {teleoperation}
\bibfield{author}{\bibinfo{person}{Jayant Singh},
  \bibinfo{person}{Aravinda~Ramakrishnan Srinivasan}, \bibinfo{person}{Gerhard
  Neumann}, {and} \bibinfo{person}{Ayse Kucukyilmaz}.}
  \bibinfo{year}{2020}\natexlab{}.
\newblock \showarticletitle{Haptic-Guided Teleoperation of a 7-DoF
  Collaborative Robot Arm With an Identical Twin Master}.
\newblock \bibinfo{journal}{\emph{IEEE Transactions on Haptics}}
  \bibinfo{volume}{13}, \bibinfo{number}{1} (\bibinfo{year}{2020}),
  \bibinfo{pages}{246--252}.
\newblock
\urldef\tempurl%
\url{https://doi.org/10.1109/TOH.2020.2971485}
\showDOI{\tempurl}


\bibitem[Slater et~al\mbox{.}(2010)]%
        {slater2010first}
\bibfield{author}{\bibinfo{person}{Mel Slater}, \bibinfo{person}{Bernhard
  Spanlang}, \bibinfo{person}{Maria~V Sanchez-Vives}, {and}
  \bibinfo{person}{Olaf Blanke}.} \bibinfo{year}{2010}\natexlab{}.
\newblock \showarticletitle{First person experience of body transfer in virtual
  reality}.
\newblock \bibinfo{journal}{\emph{PloS one}} \bibinfo{volume}{5},
  \bibinfo{number}{5} (\bibinfo{year}{2010}), \bibinfo{pages}{e10564}.
\newblock


\bibitem[Sullivan et~al\mbox{.}(2021)]%
        {sullivan2021haptic}
\bibfield{author}{\bibinfo{person}{Jennifer~L Sullivan},
  \bibinfo{person}{Shivam Pandey}, \bibinfo{person}{Michael~D Byrne}, {and}
  \bibinfo{person}{Marcia~K O'Malley}.} \bibinfo{year}{2021}\natexlab{}.
\newblock \showarticletitle{Haptic feedback based on movement smoothness
  improves performance in a perceptual-motor task}.
\newblock \bibinfo{journal}{\emph{IEEE Transactions on Haptics}}
  \bibinfo{volume}{15}, \bibinfo{number}{2} (\bibinfo{year}{2021}),
  \bibinfo{pages}{382--391}.
\newblock


\bibitem[Systems(2025)]%
        {phantom}
\bibfield{author}{\bibinfo{person}{3D Systems}.}
  \bibinfo{year}{2025}\natexlab{}.
\newblock \bibinfo{title}{PHANTOM Haptic Devices}.
\newblock
\newblock
\urldef\tempurl%
\url{https://fr.3dsystems.com/haptics-devices/3d-systems-phantom-premium}
\showURL{%
\tempurl}
\newblock
\shownote{Consulted may 12th 2025}.


\bibitem[Terenti and Vatavu(2023)]%
        {terenti2023vireo}
\bibfield{author}{\bibinfo{person}{Mihail Terenti} {and}
  \bibinfo{person}{Radu-Daniel Vatavu}.} \bibinfo{year}{2023}\natexlab{}.
\newblock \showarticletitle{Vireo: Web-based graphical authoring of
  vibrotactile feedback for interactions with mobile and wearable devices}.
\newblock \bibinfo{journal}{\emph{International Journal of Human--Computer
  Interaction}} \bibinfo{volume}{39}, \bibinfo{number}{20}
  (\bibinfo{year}{2023}), \bibinfo{pages}{4162--4180}.
\newblock


\bibitem[Weber et~al\mbox{.}(2011)]%
        {weber2011evaluation}
\bibfield{author}{\bibinfo{person}{Bernhard Weber}, \bibinfo{person}{Simon
  Sch{\"a}tzle}, \bibinfo{person}{Thomas Hulin}, \bibinfo{person}{Carsten
  Preusche}, {and} \bibinfo{person}{Barbara Deml}.}
  \bibinfo{year}{2011}\natexlab{}.
\newblock \showarticletitle{Evaluation of a vibrotactile feedback device for
  spatial guidance}. In \bibinfo{booktitle}{\emph{2011 IEEE World Haptics
  Conference}}. IEEE, \bibinfo{pages}{349--354}.
\newblock


\bibitem[WYVRN(2025)]%
        {Interhaptics}
\bibfield{author}{\bibinfo{person}{WYVRN}.} \bibinfo{year}{2025}\natexlab{}.
\newblock \bibinfo{title}{Interhaptics SDK}.
\newblock
\newblock
\urldef\tempurl%
\url{https://www.wyvrn.com/interhaptics/}
\showURL{%
\tempurl}
\newblock
\shownote{Consulté le 12 mai 2025}.


\bibitem[Yun et~al\mbox{.}(2023)]%
        {Yun2023}
\bibfield{author}{\bibinfo{person}{Gyeore Yun}, \bibinfo{person}{Minjae Mun},
  \bibinfo{person}{Jungeun Lee}, \bibinfo{person}{Dong-Geun Kim},
  \bibinfo{person}{Hong~Z Tan}, {and} \bibinfo{person}{Seungmoon Choi}.}
  \bibinfo{year}{2023}\natexlab{}.
\newblock \showarticletitle{Generating Real-Time, Selective, and Multimodal
  Haptic Effects from Sound for Gaming Experience Enhancement}. In
  \bibinfo{booktitle}{\emph{Proceedings of the 2023 CHI Conference on Human
  Factors in Computing Systems}} (Hamburg, Germany) \emph{(\bibinfo{series}{CHI
  '23})}. \bibinfo{publisher}{Association for Computing Machinery},
  \bibinfo{address}{New York, NY, USA}, Article \bibinfo{articleno}{315},
  \bibinfo{numpages}{17}~pages.
\newblock
\showISBNx{9781450394215}
\urldef\tempurl%
\url{https://doi.org/10.1145/3544548.3580787}
\showDOI{\tempurl}


\end{thebibliography}

\end{document}